\documentclass{aa}
\usepackage{graphicx}
\usepackage{txfonts}
\begin{document}
   \title{Efficiency of the centrifugally induced curvature drift instability in AGN winds.}

   \author{Osmanov Z.
          }


   \institute{E. Kharadze Georgian National Astrophysical Observatory,
              Kazbegi str. 2a, 0106 Tbilisi\\
              \email{z.osmanov@astro-ge.org}
             }



  \abstract
  {}
   {For studying how the field lines are twisting nearby the light
    cylinder surface, which provides the free motion of AGN winds through the
    mentioned area, the investigation of the centrifugally driven curvature drift instability
    is presented.
}
   {Studying the dynamics of the relativistic MHD flow close to the light
    cylinder surface, by applying a single particle approach
    based on the centrifugal acceleration,
    the dispersion relation of the instability is
    derived and analytically solved. }
   {Considering the typical values
   of AGN winds, it is shown that the time scale of the curvature drift instability
   is much less than the accretion process time scale, indicating that
   the present instability is very efficient and might strongly
   influence processes in AGN plasmas.}
   {}


   \maketitle
%


\section{Introduction}

For studying the AGN winds the fundamental problem relates to the
understanding of a question: how the plasma goes through the Light
Cylinder Surface (LCS - the hypothetical zone, where the linear
velocity of rotation equals the speed of light). An innermost
region of AGNs is characterized by the rotational motion, and it
is obvious that such a character of motion must affect the plasma
dynamics. According to the standard model, the magnetic field due
to the frozen-in condition undergoes plasmas and consequently the
flow follows the field lines, co-rotating with them. This means
that the plasma particles moving along quasi straight magnetic
field lines in the nearby area of the LCS have to reach the speed
of light. On the other hand no physical system can maintain such a
motion and this fact must result in a certain twisting process of
the magnetic field lines on the LCS. Generalizing the work:
(\cite{mr}) for curved trajectories, by \cite{r03} the dynamics of
a single particle moving along a prescribed rotating curved
channel has been investigated. It was shown that if the
trajectories are given by the Archimedes spiral, the particles can
cross the LCS avoiding the light cylinder problem. Making one more
step to this investigation is to find an appropriate mechanism,
which might provide the twisting of the magnetic field lines,
giving rise to the shape of the Archimedes spiral, which in turn
makes the dynamics force-free.


The light cylinder problem in the context of the force-free regime
has been extensively studied numerically for pulsars. The
investigation developed in (\cite{spit1,spit2}) has shown that the
plasmas can go through the LCS. This work was based on a current
generated by the electric drift (\cite{bland}), which is vanishing
for quasi neutral plasmas and can not give a contribution in the
dynamics of astrophysical flows having almost equal numbers of
positive and negative charges.

Since the innermost region of AGNs is rotating, a role of the
Centrifugal Force (CF) seems to be particularly interesting for
the study of the relativistic plasma motion. The centrifugally
driven outflows have been discussed in series of works. A special
attention deserves the work of \cite{bp82}, where the authors
discuss the possibility of the energy and angular momentum pumping
from the accretion disk, emphasizing the role of the centrifugal
acceleration in this process. By \cite{gl97} the CF was considered
in the context of the non thermal radiation from the spinning
AGNs. Generalizing the mentioned work it has been shown
(\cite{osm7,ra8}) that due to the centrifugal acceleration,
electrons gain very high energies with Lorentz factors up to
$\gamma\sim 10^8$. This means that the energy budget in the AGN
winds is very high and if one finds a mechanism of the conversion
of at least a small fraction of this energy into a variety
instabilities, one might have interesting consequences in the
physics of AGN outflows.

The centrifugal force may drive different kinds of instabilities.
Obviously the CF acting on a moving particle changes in time and
in the context of instabilities plays a role of the parameter.
Consequently the corresponding instability is called the
parametric instability.

The centrifugally driven parametric instability first has been
introduced in (\cite{incr1}) for the Crab pulsar magnetosphere. We
have argued that the centrifugal force may cause the separation of
charges, leading to the creation of an unstable electrostatic
field. Estimating the linear growth rate it has been shown that
the instability was extremely efficient. The method developed in
(\cite{incr1}) was applied for AGN jets (\cite{incr3}) for
studying the stability problem of the rotation induced
electrostatic instability and for understanding how efficient is
the centrifugal acceleration in this process. Another kind of the
instability which might be induced by the CF is the so called
Curvature Drift Instability (CDI). Even if the field lines
initially have a very small curvature, it might cause a drifting
process of plasmas, leading to the CDI. By \cite{mnras} the two
component relativistic plasma has been considered for studying the
role of the centrifugal acceleration in the curvature drift
instability for pulsar magnetospheres. The investigation has shown
that the growth rate was more than pulsar spin down rates by many
orders of magnitude, indicating high efficiency of the CDI. The
curvature drift current produces the toroidal component of the
magnetic field, which due to the efficient unstable character of
the process amplifies rapidly, changing the overall configuration
of the magnetic field. This leads to the transformation of field
lines into the shape of the Archimedes spiral, when the motion of
the particles switches to the so called force-free regime
(\cite{ff}) and the plasma goes through the LCS.

In the present paper in order to investigate the twisting process
of magnetic field lines due to the CDI, the method developed in
(Osmanov et al. 2008a, \cite{ff}) will be implemented for AGN
winds.

The paper is arranged as follows. In \S\ref{sec:consid} we
introduce the curvature drift waves and derive the dispersion
relation. In \S\ref{sec:results} the results for typical AGNs are
present and in \S\ref{sec:summary} we summarize our results.
\begin{figure}
  \resizebox{\hsize}{!}{\includegraphics[angle=0]{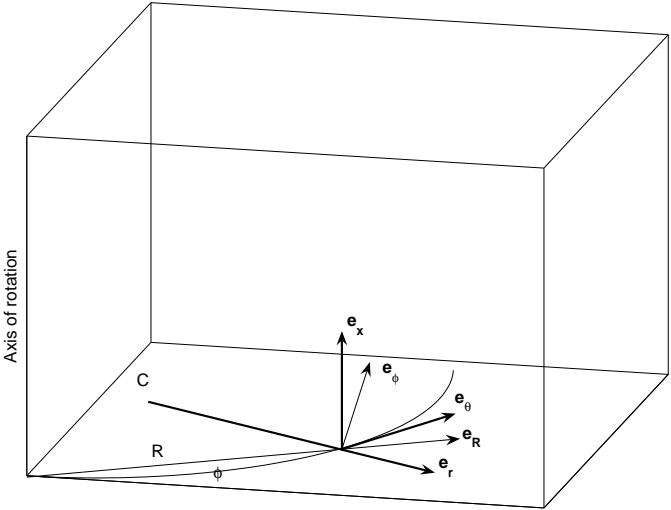}}
  \caption{Two orthonormal bases are
considered: i) cylindrical components of unit vectors, (${\bf
e}_\Phi, {\bf e}_R$, ${\bf e}_x$); ii) unit vectors of the system
rigidly fixed on each point of the curve, (${\bf e}_r, {\bf
e}_{\theta}$, ${\bf e}_x$), respectively. $C$ is the center of the
curvature. Hereafter the set of coordinates I call the field line
coordinates.}\label{fig}
\end{figure}
\section{Main consideration} \label{sec:consid}
%
%
%
We start the investigation by considering the two component plasma
consisting of the relativistic electrons with the Lorentz factor
$\gamma_{e}\sim 10^{5-8}$ (\cite{osm7,ra8}) and the bulk component
(protons) with $\gamma_{b}\sim 10$. Since we are interested in the
twisting process, we suppose that initially the field lines are
almost rectilinear in order to study how this configuration
changes in time.

Let us start by the Euler equation governing the dynamics of
plasma particles, co-rotating with the straight magnetic field
lines. By applying the method developed by \cite{chedia} one can
show that the Euler equation gets the following form:

\begin{equation}
\label{eul} \frac{\partial{\bf p_{\alpha}}}{\partial t}+({\bf
v_{\alpha}\nabla)p_{\alpha}}=
-\gamma_{\alpha}\xi{\bf\nabla}\xi+\frac{q_{\alpha}}{m_{\alpha}}\left(\bf
E+ \bf v_{\alpha}\times\bf B\right),
\end{equation}

$$\alpha=\{e,b\},$$
where $$ \label{xi} \xi\equiv \sqrt{1-\Omega^2R^2/c^2}. $$

Here ${\bf p_{\alpha}}$ is the momentum, ${\bf v_{\alpha}}$ - the
velocity and ${\gamma_{\alpha}}$ - the Lorentz factor of the
relativistic particles. ${\bf E}$ and ${\bf B}$ are the electric
field and the magnetic induction respectively. By $q_{\alpha}$ and
$m_{\alpha}$ we denote particle's charge and the rest mass
respectively. We express the equation of motion in the cylindrical
coordinates (see Fig. \ref{fig}). The first term of the right hand
side of the Euler equation $-\gamma_{\alpha}\xi{\bf\nabla}\xi$
represents the centrifugal force, which on the light cylinder
surface becomes infinity. This means that its overall effect is
significant in the nearby zone of the LCS. For describing our
physical system, one needs the full set of equations, and in order
to close the system, we add to Eq. (\ref{eul}) the continuity
equation:

\begin{equation}
\label{cont} \frac{\partial n_{\alpha}}{\partial t}+{\bf
\nabla}(n_{\alpha}{\bf v_{\alpha}})=0,
\end{equation}
and the induction equation:
\begin{equation}
\label{ind} {\bf \nabla\times B} = \frac{1}{c}\frac{\partial {\bf
E}}{\partial t}+\frac{4\pi}{c}\sum_{\alpha=e,b}{\bf J_{\alpha}},
\end{equation}
where $n_{\alpha}$ and ${\bf J_{\alpha}}$ are the density and the
current, respectively.

In the zeroth approximation the plasma particles undergo only the
centrifugal force. Different species at different positions
experience different CFs, which will cause the separation of
charges, leading to the creation of the additional electromagnetic
field considered as the first order term in our equations.

The leading state is characterized by the frozen-in condition,
$\bf E+ \bf v_{\alpha 0}\times\bf B_0=0$, which reduces Eq.
(\ref{eul}) into the following form (\cite{mr}):

\begin{equation}
\label{eul_0} \frac{dv}{dt}=\frac{\Omega^2R}{1-
\frac{\Omega^2R^2}{c^2}}\left[1-\frac{\Omega^2R^2}{c^2}-
\frac{2v^2}{c^2}\right].
\end{equation}

The present equation of motion describing the kinematic behaviour
of the single co-rotating particle has the following solution:

\begin{equation}
\label{v} v(t)\equiv v_{_\parallel} \approx c\cos(\Omega t),
\end{equation}
for ultra relativistic cases ($\gamma>>1$) and the following
initial conditions: $R(0) = 0$, $v(0)\approx c$ ($v\equiv dR/dt)$.
Here $v_{_\parallel}$ denotes the velocity component along the
magnetic field lines.

Since we suppose that the magnetic field lines initially have the
small curvature, the particles moving radially, will drift along
the $x$ axis as well (Osmanov et al. 2008a) (see Fig. \ref{fig}).
The mentioned drift of charges will produce the corresponding
current, which inevitably will create the toroidal magnetic field,
changing the overall configuration of the field lines. Therefore
the aim of the present work is to study the role of the
centrifugally induced curvature drift instability in the twisting
process of the magnetic field lines. For this purpose one can
linearize the system of equations Eqs. (\ref{eul}-\ref{ind}),
perturbing all physical quantities around the leading state:

\begin{equation}
\label{expansion} \Psi\approx \Psi^0 + \Psi^1,
\end{equation}

\begin{equation}
\Psi = \{n,{\bf v},{\bf p},{\bf E},{\bf B}\}.\end{equation}
Let us express the perturbation by following:

\begin{equation}
\label{pert} \Psi^1(t,{\bf r})\propto\Psi^1(t)
\exp\left[i\left({\bf kr} \right)\right] \,,
\end{equation}
then from Eqs. (\ref{eul}-\ref{ind}) one can derive the linearized
set of equations governing the CDI:

\begin{equation}
\label{eulp} \frac{\partial p^1_{{\alpha}x}}{\partial
t}-i(k_xu_{\alpha}+k_{\phi}v_{_\parallel})p^1_{{\alpha}x}=
\frac{q_{\alpha}}{m_{\alpha}}v_{_\parallel}B^1_{r},
\end{equation}
\begin{equation}
\label{contp} \frac{\partial n^1_{\alpha}}{\partial
t}-i(k_xu_{\alpha}+k_{\phi}v_{_\parallel})n^1_{\alpha}=
ik_xn_{\alpha}^0v^1_{\alpha x},
\end{equation}
\begin{equation}
\label{indp} -ik_{\phi}cB^1_{r} = 4\pi
\sum_{\alpha=e,b}q_{\alpha}(n_{\alpha}^0v^1_{\alpha
x}+n_{\alpha}^1u_{\alpha}).
\end{equation}
Here by $u_{\alpha}$ we denote the curvature drift velocity along
the $x$ axis:

\begin{equation}
\label{drift} u_{\alpha}= \frac{\gamma_{\alpha_0}
v_{_\parallel}^2}{\omega_B R_B},
\end{equation}
where $\omega_{\alpha B} = q_{\alpha}B_0/m_{\alpha}c$; $R_B$ is
the curvature radius of magnetic field lines; and

\begin{equation}
\label{equipart}B_0 = \sqrt{\frac{2L}{R_{lc}c^2}}
\end{equation}
is the equipartition magnetic induction on the LCS for the leading
state (here $L$ is the luminosity of the AGN and $R_{lc}=c/\Omega$
is the light cylinder radius). For deriving Eqs.
(\ref{eulp}-\ref{indp}) the wave propagating almost perpendicular
to the equatorial plane has been considered and the expression:
$v^1_r\approx cE^1_x/B_{0}$ was taken into account. For simplicity
the set of equations are given in the coordinates of the field
line (see Fig. \ref{fig}).

Let us express $v^1_{\alpha x}$ and $n^1_{\alpha}$ by the
following way:

\begin{equation}
\label{anzp} v^1_{\alpha x}\equiv V_{\alpha x}e^{i{\bf
kA_{\alpha}(t)}},
\end{equation}
\begin{equation}
\label{anzn} n^1_{\alpha}\equiv N_{\alpha}e^{i{\bf
kA_{\alpha}}(t)},
\end{equation}
where
\begin{equation}
\label{Ax} A_{\alpha x}(t) = \frac{u_{\alpha}t}{2} +
\frac{u_{\alpha}}{4\Omega}\sin(2\Omega t),
\end{equation}
\begin{equation}
\label{Af} A_{\alpha\phi}(t) = \frac{c}{\Omega}\sin(\Omega t).
\end{equation}
Then, by substituting Eqs. (\ref{anzp},\ref{anzn}) into Eqs.
(\ref{eulp}-\ref{indp}), it is easy to solve the system for the
toroidal component:

$$ -ik_{\phi}cB^1_{r}(t)
=\sum_{\alpha=e,b}\frac{\omega^2_{\alpha}}{\gamma_{\alpha_0}}{\rm
e}^{i{\bf kA_{\alpha}}(t)}\int^t{\rm e}^{-i{\bf
kA_{\alpha}}(t')}v_{_\parallel}(t')B_{r}(t')dt'+ $$
$$i\sum_{\alpha=e,b}\frac{\omega^2_{\alpha}}{\gamma_{\alpha_0}}k_xu_{\alpha}{\rm
e}^{i{\bf kA_{\alpha}}(t)}\int^tdt'\int^{t''}{\rm e}^{-i{\bf
kA_{\alpha}}(t'')}v_{_\parallel}(t'')B_{r}(t'')dt''$$
\begin{equation}
\label{ind1}
\end{equation}
where $\omega_{\alpha} = e\sqrt{4\pi n_{\alpha}^0/m_{\alpha}}$ is
the plasma frequency.  After making the Fourier transform (see
Appendix A), Eq. (\ref{ind1}) gets the form:

$$B_{r}(\omega) =
-\sum_{\alpha=e,b}\frac{\omega^2_{\alpha}}{2\gamma_{\alpha_0}k_{\phi}c}\sum_{\sigma
= \pm
1}\sum_{s,n,l,p}\frac{J_s(g_{\alpha})J_n(h)J_l(g_{\alpha})J_p(h)}{\omega
+ \frac{k_xu_{\alpha}}{2}+\Omega (2s+n) } \times$$ $$\times
B_{r}\left(\omega+\Omega
\left(2[s-l]+n-p+\sigma\right)\right)\left[1-\frac{k_xu_{\alpha}}{\omega
+ \frac{k_xu_{\alpha}}{2}+\Omega (2s+n)}\right]$$
$$+\sum_{\alpha=e,b}\frac{\omega^2_{\alpha}k_xu_{\alpha}}{4\gamma_{\alpha_0}k_{\phi}c}\sum_{\sigma,\mu
= \pm
1}\sum_{s,n,l,p}\frac{J_s(g_{\alpha})J_n(h)J_l(g_{\alpha})J_p(h)}{\left(\omega
+ \frac{k_xu_{\alpha}}{2}+\Omega (2[s+\mu]+n)\right)^2 } \times$$
\begin{equation}
\label{disp1} \times B_{r}\left(\omega+\Omega
\left(2[s-l+\mu]+n-p+\sigma\right)\right),
\end{equation}
where $$g_{\alpha} = \frac{k_xu_{\alpha}}{4\Omega},
\;\;\;\;\;\;\;\;\;\;\;\;h = \frac{k_{\phi}c}{\Omega}.$$

Let us note that Eq. (\ref{disp1}) is written for $B_r(\omega)$
and on the right hand side of the equation, there is the infinite
number of components with $B_r(\omega\pm\Omega)$, $B_r(\omega\pm
2\Omega)$,... etc. This means that for solving the mentioned
equation, thus for closing the system, one needs to add the
corresponding expressions for $B_r(\omega\pm\Omega)$,
$B_r(\omega\pm 2\Omega)$,... etc. But then the system becomes
composed of the infinite number of equations, making the task
unsolvable. In order to overcome this problem one has to use a
certain, physically reasonable cutoff on the right hand side of
the equation (\cite{silin}). As it is clear from Eq.
(\ref{disp1}), the considered instability is characterized by the
following proper frequency of the curvature drift modes:

\begin{equation}
\label{freq} \omega_0\approx -\frac{k_xu_{\alpha}}{2},
\end{equation}
when the corresponding conditions: $k_xu_{\alpha}/2<0$, $2s+n = 0$
and $2[s+\mu]+n = 0$ are satisfied. As we will see only the
resonance terms give the significant contribution to the result.

Let us consider parameters $L\sim 10^{44}erg/s$, $\Omega=3\times
10^{-5}s^{-1}$, $\gamma_{e0}\sim 10^5$, $R_B\approx R_{lc}$
$n_{e0}\sim 0.001cm^{-3}$ typical for AGN winds. Then examining
the curvature drift waves with $\lambda\sim R_{lc}$
($\lambda=2\pi/k$ is the wave length), one can show that $|k_x
u_{e}/2|\sim 10^{-12}s^{-1}<<\Omega$ (here, it is supposed that
$k_x<0$ and $u_{e0}>0$, otherwise the frequency becomes negative).
Therefore all terms with non zero $\Omega (2s+n)$ and $\Omega
(2[s+\mu]+n)$ are rapidly oscillative and do not contribute in the
final result. Consequently the only terms, which influence the
solution of Eq. (\ref{disp1}) are the leading terms, the
contribution of which reduces the equation (Osmanov et al.2008a)
(see Appendix B):

$$\left(\omega + \frac{k_xu_{e}}{2}\right)^2 \approx
\sum_{\sigma,\mu = \pm
1}\sum_{s,l}\Xi_{\mu}J_s(g)J_{n'(s,\mu)}(h)J_l(g)J_{p'(l,\sigma)}(h),$$
\begin{equation}
\label{disp}
\end{equation}
where
\begin{equation}
\label{xi}\Xi_0 = 2\Xi_{\pm 1} =
\frac{\omega^2_{e}k_xu_{e}}{2\gamma_{ep_0}k_{\theta}c},
\end{equation}
$$n' = -2(s+\mu),\;\;\;\;\;\;\; p'=-2l+\sigma,\;\;\;\;\;\;\;
g\equiv g_{e}.$$
Note, that Eq. (\ref{disp}) does not consist of the bulk
components at all, because as a direct calculation shows, their
contribution is negligible with respect to the terms corresponding
to the relativistic electrons. Expressing the frequency by the
real and imaginary parts: $\omega\equiv\omega_0+i{\Gamma}$ it is
straightforward to show from Eq. (\ref{disp}), that the increment
of the CDI writes as :
\begin{equation}
\label{increm} \Gamma\approx \left[\sum_{\sigma,\mu = \pm
1}\sum_{s,l}\Xi_{\mu}J_s(g)J_{-2(s +
\mu)}(h)J_l(g)J_{-2l+\sigma}(h)\right]^{\frac{1}{2}}.
\end{equation}

\begin{figure}
  \resizebox{\hsize}{!}{\includegraphics[angle=0]{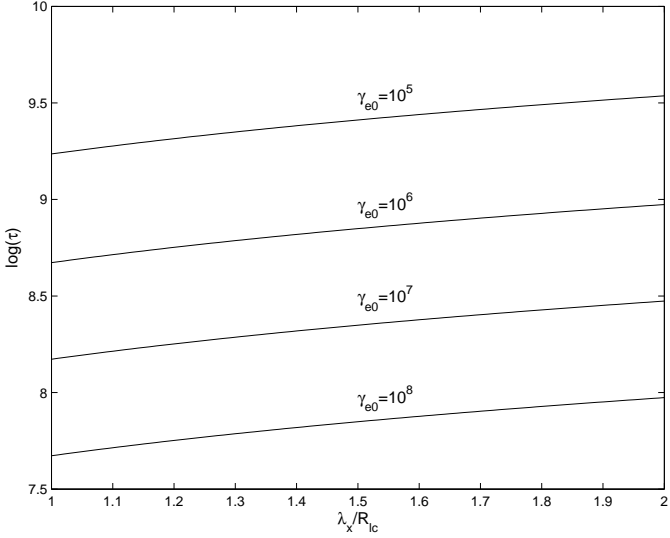}}
  \caption{The dependence of logarithm of the instability time scale
  on the normalized wave length. The set of parameters is:
  $\gamma_{e0} = \{10^5;10^6;10^7;10^8\}$, $R_B\approx R_{lc}$, $n_{e0} =
0.001cm^{-3}$, $\lambda_{\phi} = 100R_{lc}$ and $L/L_E =
0.01$.}\label{lambda}
\end{figure}
\begin{figure}
  \resizebox{\hsize}{!}{\includegraphics[angle=0]{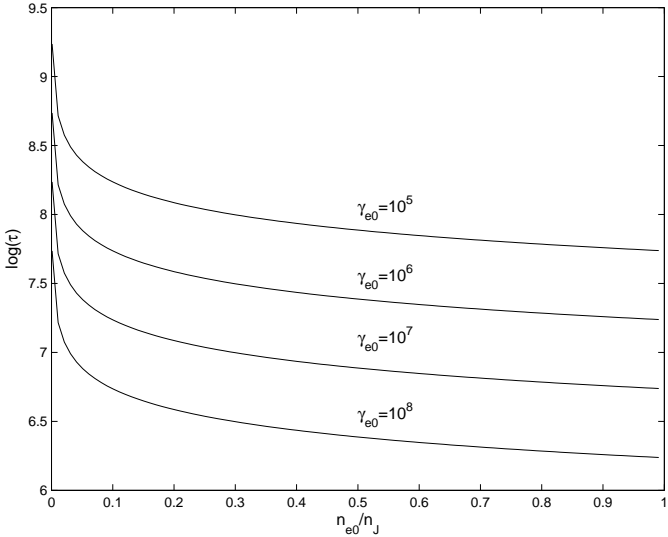}}
  \caption{The dependence of logarithm of the instability time scale
  on the density normalized by the medium density. The set of parameters is:
  $\gamma_{e0} = \{10^5;10^6;10^7;10^8\}$, $R_B\approx R_{lc}$, $\lambda_{e0} =
R_{lc}$, $\lambda_{\phi} = 100R_{lc}$ and $L/L_E =
0.01$.}\label{density}
\end{figure}

\section{Results} \label{sec:results}
%
%
%
In this section we investigate the efficiency of the CDI for AGN
winds. The behaviour of the growth rate will be studied versus the
wave length, the density of relativistic electrons, their Lorentz
factors and the AGN bolometric luminosity.

For studying the behaviour of the instability as a function of the
wave length one can examine the typical AGN parameters:
$M_{BH}=10^{8}\times M_{\odot}$, $\Omega = 5\times 10^{-5}s^{-1}$
and $L = 10^{44}erg/s$, where $M_{BH}$ is the AGN mass,
$M_{\odot}$ - solar mass and $L$ is the bolometric luminosity of
the AGN.

Let us consider Eq. (\ref{increm}) and plot logarithm of the
instability time scale

\begin{equation}
\label{tau} \tau\equiv \frac{1}{\Gamma}
\end{equation}
versus the wave length normalized by the light cylinder radius.
The present consideration is based on the centrifugal
acceleration. As it has been shown in (\cite{osm7}), due to the
CF, the relativistic particles may reach very high Lorentz
factors. For this purpose it is reasonable to investigate the
efficiency of the instability versus the wave length but for
different values of Lorentz factors. Fig. \ref{lambda} shows the
mentioned behaviour for the following parameters: $\gamma_{e0} =
\{10^5;10^6;10^7;10^8\}$, $R_B\approx R_{lc}$, $n_{e0} =
0.001cm^{-3}$, $\lambda_{\phi}\equiv 2\pi/k_{\phi} = 100R_{lc}$
and $L/L_E = 0.01$, where $L_E = 10^{46}erg/s$ is the Eddington
luminosity for the given AGN mass. Different curves correspond to
different values of Lorentz factors. As it is clear from the
plots, the time scale is a continuously increasing function of
$\lambda_x(\equiv 2\pi/k_x)$, which is a direct consequence of
Eqs. (\ref{xi},\ref{increm}). Indeed, as we see from Eq.
(\ref{xi}), $\Xi_{0,\pm 1}\sim 1/{\lambda_x}$, which combining
with Eq. (\ref{increm}) gives $\Gamma\sim 1/\sqrt{\lambda_x}$.
Therefore the bigger the initial perturbation wave length, the
less the instability time scale and consequently the less the CDI
efficiency. For the given range of $\lambda_x$ and different
values of $\gamma_{e0}$, the CDI time scale varies from $\sim
10^7s$ ($\lambda_{x}/R_{lc} = 1$, $\gamma_{e0} = 10^8$) to $\sim
10^9$ ($\lambda_{x}/R_{lc} = 2$, $\gamma_{e0} = 10^5$).

Sine the CDI growth rate depends on the plasma frequency [see Eq.
(\ref{increm},\ref{xi})], which in turn is the function of the
density, it is obvious that the instability time scale must be
influenced by the density of relativistic electrons in the AGN
winds. Indeed, In Fig. \ref{density} the plots of $log(\tau)$
versus the AGN wind density are shown and one can see that the
time scale is the continuously decreasing function of
$n_{e0}/n_m$. The set of parameters is the same as in Fig.
\ref{lambda}, except $\lambda_{x} = R_{lc}$ and
$n_{e0}/n_m\in\{0.001,1\}$. Here $n_{e0}$ is normalized by the
intergalactic medium density, $n_m\approx 1cm^{-3}$. As we see
from the figure, $\tau$ varies from $\sim 10^9s$ ($n_{e0}/n_m =
0.001$, $\gamma_{e0} = 10^5$) to $\sim 10^6s$ ($n_{e0}/n_m = 1$,
$\gamma_{e0} = 10^8$).


\begin{figure}
  \resizebox{\hsize}{!}{\includegraphics[angle=0]{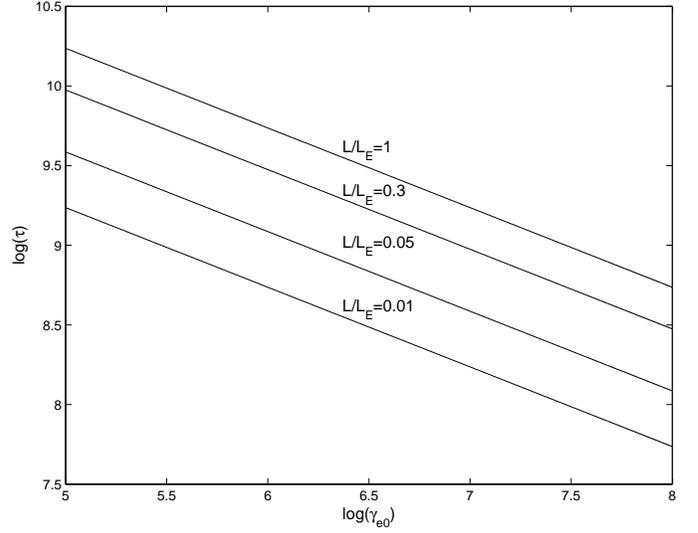}}
  \caption{The dependence of logarithm of the instability time scale
  on $log(\gamma_{e0})$. The set of parameters is:
  $R_B\approx R_{lc}$, $n_{e0}/n_m = 0.001$, $\lambda_{e0} =
R_{lc}$, $\lambda_{\phi} = 100R_{lc}$ and $L/L_E =
\{0.01;0.05;0.3;1\}$.}\label{gamma}
\end{figure}

In Fig. \ref{gamma} the plot of $log(\tau)$ versus
$log(\gamma_{e0})$ is shown for different luminosities. The set of
parameters is the same as in Fig. \ref{lambda} except the
continuous range of $\gamma_{e0}\in\{10^5;10^8\}$, different
values of the luminosity $L/L_E=\{0.01;0.05;0.3;1\}$ and
$\lambda_{x} = R_{lc}$. The figure shows the continuously
decreasing behaviour of the instability time scale, which is a
natural consequence of the fact that the more energetic electrons
will induce the curvature drift instability more efficiently.
Indeed, from Eq. (\ref{drift}) it is clear that the drift velocity
is proportional to the Lorentz factor of the particle, and hence
the corresponding instability will be more efficient, leading to
the decreasing behaviour of $log(\tau)$. As we see the instability
time scale varies from $\sim 10^{10}s$ ($\gamma_{e0} = 10^5$,
$L/L_E=1$) to $\sim 10^8s$ ($\gamma_{e0} = 10^7$, $L/L_E=0.01$).
On the other hand the plots for different luminosities exhibit
another feature of the behaviour of $\tau$: by increasing the
luminosity of the AGN, the corresponding instability becomes less
efficient.

\begin{figure}
  \resizebox{\hsize}{!}{\includegraphics[angle=0]{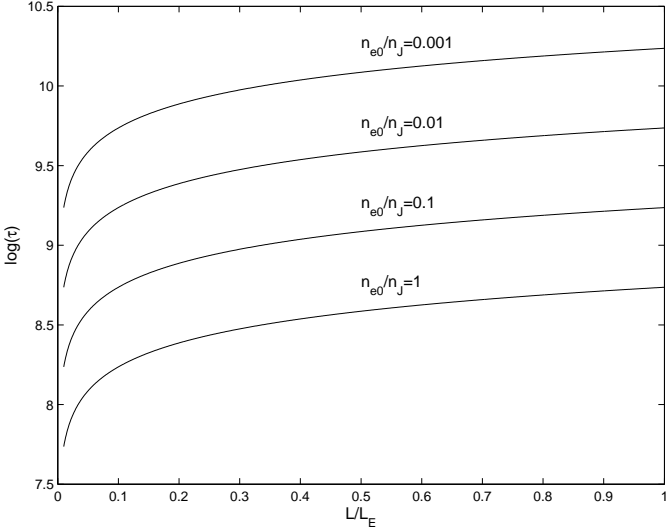}}
  \caption{The dependence of logarithm of the instability time scale
  on $log(\gamma_{e0})$. The set of parameters is: $R_B\approx
  R_{lc}$, $n_{e0}/n_m = \{0.001;0.01;0.1;1\}$, $\lambda_{e0} =
R_{lc}$, $\lambda_{\phi} = 100R_{lc}$ and $L/L_E =
0.01$.}\label{lumin}
\end{figure}

To see this particular feature more clearly, let us look at Fig.
\ref{lumin} exhibiting the dependence of $log(\tau)$ on $L/L_E$
for different values of densities. From the plots it is seen that
by increasing the luminosity, the time scale continuously
increases. This behaviour follows from the fact, that the bigger
the luminosity, the bigger the magnetic field [see Eq.
(\ref{equipart})] and hence the less the drift velocity, leading
to the less efficient CDI shown in the figure. By considering
bigger values of densities, the CDI becomes more efficient, which
we have already explained while considering Fig. \ref{density}.
For the mentioned area of quantities (see Fig. \ref{lumin}), the
time scale varies from $\sim 10^7s$ ($L/L_E = 0.01$, $n_{e0}/n_J =
1$) to $\sim 10^{10}s$ ($L/L_E = 1$, $n_{e0}/n_J = 0.001$).

We see from the present investigation that the instability time
scale varies in the following range: $\tau\in\{10^6;10^{10}\}s$.
In order to specify how efficient is the CDI it is sensible to
examine an accretion process, estimating the corresponding
evolution time scale, and compare it with that of the CDI.

Considering the problem of fuelling of AGNs, in (\cite{king}) it
has been shown that the self gravitating mass in accretion flows
can be estimated by the following expression:

$$M_{sg}  =2.76\times
10^5\left(\frac{\eta}{0.03}\right)^{-2/27}\left(\frac{\epsilon}{0.1}\right)^{-5/27}\left(\frac{L}{0.1L_E}\right)^{5/27}
$$
\begin{equation}
\label{msg}\times
\left(\frac{M_{BH}}{10^8M_{\odot}}\right)^{23/27}M_{\odot},
\end{equation}
where $\eta$ and $\epsilon$ are Shakura, Sunyaev viscosity
parameter (\cite{sak}) and the accretion parameter respectively.
The latter can be defined by means of the accretion mass rate
$\dot{M}$ and the luminosity:

\begin{equation}
\label{epsilon}\epsilon\equiv \frac{L}{\dot{M}c^2}.
\end{equation}
Then, defining the accretion time scale $t_{evol}\equiv
M_{sg}/\dot{M}$ one can reduce it to (\cite{king}):

$$t_{evol}  =3.53\times
10^{13}\left(\frac{\eta}{0.03}\right)^{-2/27}\left(\frac{\epsilon}{0.1}\right)^{22/27}\left(\frac{L}{0.1L_E}\right)^{-22/27}
$$
\begin{equation}
\label{tev}\times
\left(\frac{M_{BH}}{10^8M_{\odot}}\right)^{-4/27}s.
\end{equation}
As it is clear from Eq. (\ref{tev}), the accretion evolution time
scale depends on two major AGN parameters: on the luminosity ($L$)
and the AGN mass ($M_{BH}$). Bearing in mind Eq. (\ref{tev}) it is
sensible to investigate $t_{evol}$ versus $L$ and $M_{BH}$. For
this purpose let us consider the possible maximum "area" of the
parametric space, $L-M_{BH}$ studying the behaviour of $t_{evol}$
for the typical accretion disk parameters $\eta = 0.03$, $\epsilon
= 0.1$.

\begin{figure}
  \resizebox{\hsize}{!}{\includegraphics[angle=0]{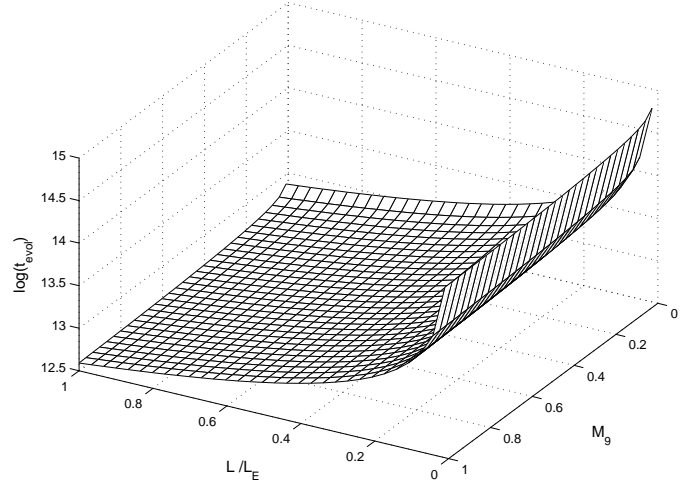}}
  \caption{The behaviour of logarithm of the evolution time scale versus $M_9$
  and $L/L_E$. The set of parameters is: $\eta = 0.03$, $\epsilon = 0.1$.}
  \label{evol}
\end{figure}

In Fig. \ref{evol} we show the two dimensional surface of
logarithm of the evolution time scale. The variables are in the
following range: $M_{9}\equiv M_{BH}/(10^9\times
M_{\odot})\in\{0.01;1\}$ and $L/L_E\in\{0.01;1\}$. For plotting
the figure we took into account that according to the observations
AGN masses vary in the following range:
$M_{BH}\in\{10^6;10^9\}\times M_{\odot}$ (\cite{agnmas}). As it is
clear from Fig. \ref{evol}, $t_{evol}$ is a continuously
decreasing function of $M_9$ and $L/L_E$. The minimum value of the
evolution time scale ($\sim 10^{12}s$), when the accretion process
is extremely efficient corresponds to $M_9 = 1$ and $L/L_E = 1$,
whereas the maximum value being of the order $\sim 10^{15}s$
corresponds to the following pair of the variables: $M_9 = 0.001$
and $L/L_E = 0.01$.

For understanding how efficient is the curvature drift
instability, one has to compare the corresponding time scale with
the evolution time scale. As it has been found, depending on
physically reasonable parameters $\tau$ varies in the range: $\sim
10^{6-10}s$, whereas the sensible area of $t_{evol}$ is: $\sim
10^{12-15}s$. Therefore the instability time scale is less than
the evolution time scale of the accretion by may orders of
magnitude, which means that the linear stage of the CDI is
extremely efficient.

The twisting process of magnetic field lines requires a certain
amount of energy and it is natural to study also the energy budget
of this process. For this reason one has to introduce the maximum
of the possible luminosity $L_{max} = \dot{M}c^2$ and compare it
with the "luminosity" corresponding to the reconstruction of the
magnetic field configuration $L_m\equiv \Delta E_m/\Delta
t\approx\Delta E_m/\tau$. (here $\Delta E_m$ is the variation of
the magnetic energy due to the curvature drift instability).

Let us consider the AGN with the luminosity, $L = 10^{45}erg/s$,
then by applying Eq. (\ref{epsilon}) and taking into account
$\epsilon = 0.1$ one can show that the accretion may provide the
following maximum value:

\begin{equation}
\label{maxlum} L_{max} = 10^{46}erg/s.
\end{equation}
On the other hand if the process of sweepback is realistic, the
magnetic "luminosity" can not exceed $L_{max}$. Since this process
takes place in the nearby zone of the LCS, the effective spatial
volume can be estimated by the following expression:

\begin{equation}
\label{volum} \Delta V\approx R_{lc}^2\Delta R=R_{lc}^3\kappa,
\end{equation}
\begin{figure}
  \resizebox{\hsize}{!}{\includegraphics[angle=0]{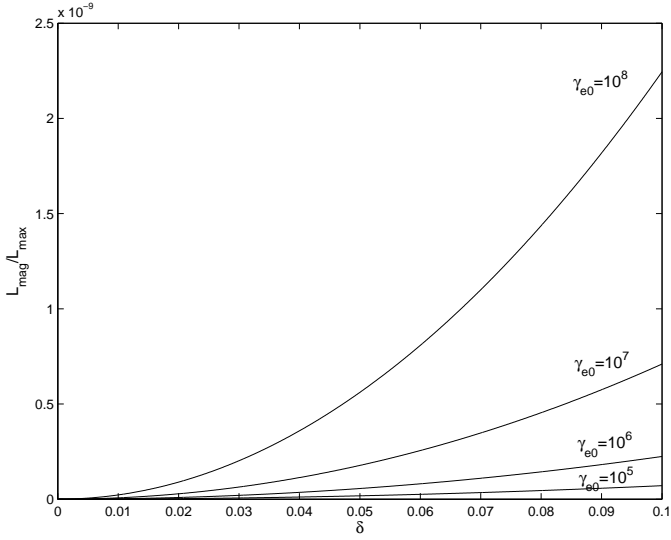}}
  \caption{The dependence of $L_{m}/L_{max}$ versus $\delta$. The set of parameters is:
  $\gamma_{e0} =
\{10^5;10^6;10^7;10^8\}$, $R_B\approx R_{lc}$, $n_{e0} =
0.001cm^{-3}$, $\lambda_{\phi} = 100R_{lc}$, $\lambda_{x} =
R_{lc}$ and $L = 10^{45}erg/s$.}
  \label{energy}
\end{figure}
where $\kappa\equiv \Delta R/R_{lc}<<1$ represents the non
dimensional thickness of the considered spatial zone. Taking into
account Eq. (\ref{volum}) the expression of the magnetic
"luminosity" reduces to:

\begin{equation}
\label{maglum} L_{m} = \frac{B_r^2}{4\pi\tau}R_{lc}^3\kappa,
\end{equation}
where $B_r$ due to the CDI behaves in time as:

\begin{equation}
\label{br} B_r = B_r^0e^{t/\tau}
\end{equation}
Here $B_r^0$ is the initial perturbation of magnetic field's
toroidal component.

Let us introduce the initial non dimensional perturbation,
$\delta$, defined as: $\delta\equiv B_r^0/B_0$, where by $B_0$ we
denote the induction of the magnetic field in the leading state
[see Eq. (\ref{br})]. By considering the following set of
parameters: $\gamma_{e0} = \{10^5;10^6;10^7;10^8\}$, $R_B\approx
R_{lc}$, $n_{e0} = 0.001cm^{-3}$, $\lambda_{\phi} = 100R_{lc}$,
$\lambda_{x} = R_{lc}$ and $L = 10^{45}erg/s$, one can plot the
behaviour of $L_{m}/L_{max}$ versus the initial perturbation for
the characteristic time scale ($t\approx\tau$). As we see from
Fig. \ref{energy}, $L_{m}/L_{max}$ varies from $\sim 0$ ($\delta =
0$) to $\sim 2.3\times 10^{-9}$ ($\delta = 0.1$, $\gamma_{e0} =
10^8$). Therefore the maximum luminosity (thus the total
luminosity budget) exceeds by many orders of magnitude the
magnetic "luminosity" required for the twisting of the field
lines. This means that only a tiny fraction of the total energy
goes to the sweepback, making this process feasible.

\section{Summary} \label{sec:summary}
%
%
%

\begin{enumerate}
      \item Considering the relativistic two component plasma for
      AGN winds the centrifugally driven
      curvature drift instability has been studied.

      \item Taking into account a quasi single approach for the particle dynamics,
      we linearized the Euler, continuity and induction equations. The dispersion
      relation characterizing the parametric instability of the toroidal component of the magnetic
      field has been derived.

      \item Considering the proper frequency of the curvature drift modes,
      the corresponding expression of the instability increment has been obtained for
      the light cylinder region.

      \item Efficiency of the CDI has been investigated
      by adopting four physical parameters, namely: the wave
      length, the flow density and the Lorentz factors of electrons and the luminosity
      of AGNs.

      \item By considering the evolution process of the accretion,
      the corresponding time scale has been estimated for a
      physically reasonable area in the parametric space
      $L-M_{BH}$. It was shown that the instability time scale was
      less by many orders of magnitude than the evolution time
      scale, indicating extremely high efficiency of the CDI.

      \item Examining the instability from the point of view of
      the energy budget, we have seen that the sweepback of the
      magnetic field lines requires only a small fraction of the
      total energy, which means that the CDI is a realistic
      process.

\end{enumerate}

      An important restriction of the present work is that for describing
      the plasma kinematics a single particle approach has been
      applied. On the other hand it is natural to suppose that the
      collective phenomena must strongly influence the overall
      kinematic picture of the plasma motion. For understanding how
      the mentioned fact changes the instability, one has to generalize the present model
      for a more realistic astrophysical scenario.

      The next limitation is related to the fact that the magnetic
      field lines were supposed to be quasi rectilinear, whereas
      in real astrophysical situations the field lines might be
      initially curved. This particular case also needs to be
      studied generalizing the present model.

      In this paper the field lines located in the equatorial plane have been
      considered, although in realistic situations the magnetic field lines
      also might be inclined with respect to the equatorial plane.
      Therefore it is essential to examine this particular case as well
      and see how the efficiency of the CDI changes, when the mentioned
      inclination angle is taken into account.

\begin{acknowledgements}
I thank professor G. Machabeli for valuable discussions. The
research was supported by the Georgian National Science Foundation
grant GNSF/ST06/4-096.\end{acknowledgements}

\end{document}